\documentclass[epj]{svjour}
\usepackage{graphics}
\usepackage{graphics}
\usepackage{graphicx}
\usepackage{subfigure}
\usepackage{amssymb}
\usepackage{CJK}
\usepackage{indentfirst}
\usepackage{amsmath}
\usepackage[square, comma, sort&compress, numbers]{natbib}

\begin{document}

\title{Testing and selecting cosmological models with ultra-compact radio quasars}

\author{Xiaolei Li\inst{1,2}, Shuo Cao\inst{1}\thanks{\emph{e-mail:} caoshuo@bnu.edu.cn}, Xiaogang Zheng\inst{1,3}, Jingzhao Qi\inst{1}, Marek Biesiada\inst{1,3}, \and Zong-Hong Zhu\inst{1}
%
}                     
%
%
\institute{Department of Astronomy, Beijing Normal University,
Beijing, 100875, China; \and Department of Physics, University of
Michigan, 450 Church Street, Ann Arbor, MI 48109, USA; \and
Department of Astrophysics and Cosmology, Institute of Phyisics,
University of Silesia, Uniwersyecka 4, 40-007, Katowice, Poland}

\date{Received: date / Revised version: date}

\abstract{ In this paper, we place constraints on four alternative
cosmological models under the assumption of the spatial flatness of
the Universe: CPL, EDE, GCG and MPC. A new compilation of 120
compact radio quasars observed by very-long-baseline interferometry,
which represents a type of new cosmological standard rulers, are
used to test these cosmological models. Our results show that the
fits on CPL obtained from the quasar sample are well consistent with
those obtained from BAO. For other cosmological models considered,
quasars provide constraints in agreement with those derived with
other standard probes at $1\sigma$ confidence level. Moreover, the
results obtained from other statistical methods including Figure of
Merit, $Om(z)$ and statefinder diagnostics indicate that: (1) Radio
quasar standard ruler could provide better statistical constraints
than BAO for all cosmological models considered, which suggests its
potential to act as a powerful complementary probe to BAO and galaxy
clusters. (2) Turning to $Om(z)$ diagnostics, CPL, GCG and EDE
models can not be distinguished from each other at the present
epoch. (3) In the framework of statefinder diagnostics, MPC and EDE
will deviate from $\rm{\Lambda}$CDM model in the near future, while
GCG model cannot be distinguished from $\rm{\Lambda}$CDM model
unless much higher precision observations are available. }
\PACS{ {cosmology}{dark energy}   \and {cosmology} {observations}
\and {methods}{statistical}
     } 
%

\authorrunning{Xiaolei Li, et al.}

\titlerunning{Probing cosmological models with QSOs}
\maketitle

\section{Introduction} \label{sec:intro}

It is strongly indicated that the Universe has entered a stage of
accelerated expansion, which was confirmed by a lot of recent
observations including Supernova Ia (SN Ia)
\citep{riess1998observational,riess2007new,suzuki2012hubble}, baryon
acoustic oscillation (BAO) \citep{percival2010baryon}, and precise
measurements of the spectrum of cosmic microwave background (CMB)
\citep{spergel2003first,komatsu2011seven,hinshaw2013nine,ade2016planck}.
However, it remains a big puzzle in modern cosmology about the
origin to the current cosmic acceleration, which gives birth to a
large class of cosmological models mathematically explaining this
phenomenon. In general, these cosmological scenarios are mainly
split into two large categories, the first of which adheres to
General Relativity and drives the current accelerated expansion
through a dark energy component, while the second focus on
gravitational modifications without the inclusion of exotic dark
energy.

In the first scenario, the most simple candidate for dark energy is
the cosmological constant $\rm{{\rm{\Lambda}}}$, in which the
equation of state (EoS) of dark energy is equal to $-1$. This model,
the so-called $\rm{\Lambda}$CDM, provides excellent agreements with
a wide range of astronomical data so far
\citep{cao2011testing,li2016comparison}. However, it is confronted
with some theoretical problems including the well-known fine tuning
problem and coincidence problem \citep{weinberg1989cosmological}.
Other models allowing any constant dark energy equation of state
(quintessence \citep{ratra1988cosmological,peebles1988cosmology},
phantom models \citep{caldwell2002phantom}, etc.), as well as models
in which the dark energy equation of state is allowed to evolve with
time (the well-known CPL parametrization
\citep{chevallier2001accelerating,linder2003cosmic}) have been
extensively studied with various astrophysical probes in the
literature \citep{cao2014cosmic}. Meanwhile, considering the
possible interaction between dark energy and dark matter, the
so-called interacting dark energy model
\citep{caldera2009dynamics,valiviita2010observational,zheng2017ultra}
could also contribute to the alleviation of the coincidence problem.
Originating from different aspects of new physics, many other
dynamical dark energy models such as the Chaplygin gas
\citep{kamenshchik2001alternative,barreiro2008wmap} and the
holographic dark energy models
\citep{huang2004holographic,setare2007holographic,sheykhi2011interacting}
have been explored by cosmologists for a long time. In the second
scenario, significant interest in modifications to general
relativity has also gained a lot attention
\citep{freese2002cardassian,Qi2017,Xu2017}, with the aim of
explaining the acceleration of the Universe without introducing dark
energy.

In the face of so many competing cosmological models, many authors
turned to various observational probes such as Supernovae
\uppercase\expandafter{\romannumeral1}a acting as standard candles
($z\sim 1.40$), strong gravitational lensing systems
\citep{cao2012constraints,cao2015cosmology}, BAO, and CMB ($z\sim
1000$) acting as standard rulers, with the aim to determine which
one is most favored by the observational data
\citep{cao2014cosmic,li2016comparison}. However, in order to achieve
this difficult goal of model filtration, it is still necessary to
acquire high precision data set and develop new complementary
techniques bridging the redshift gap of current data. In the past
decades, different types of radio sources have been proposed as
possible candidates for standard rulers in cosmological studies
\citep{buchalter1998constraining,gurvits1998angular,daly2003model}.
For instance, the size of the line emitting regions was employed as
a standard ruler to study the local expansion history of the
Universe, using which Ref.~\citep{watson2011new} actually derived
the distance to Active Galactic Nuclei (AGN). Another new type of
useful cosmological ruler is Super-Eddington accreting quasar, the
properties of which have been extensively studied in
Refs.~\cite{wang2013super,marziani2014highly}. Recent studies have
used the nonlinear relation between the ultraviolet and X-ray
luminosity of quasars to place constraints on cosmological
parameters \citep{risaliti2015hubble}. In this paper, we highlight
the usefulness of ultra-compact structure in radio quasars as a
reliable cosmic standard probe to assess some popular cosmological
models. In the more recent work \citep{cao2017a}, a sub-sample of
120 intermediate-luminosity quasars in the redshift range of
$0.46<z<2.8$ was extracted from 613 milliarcsecond ultra-compact
radio sources observed by the very-long-baseline interferometry
(VLBI) all-sky survey. A pioneer work using this data was also
performed in Ref.~\citep{cao2017ultra} to probe a flat
$\rm{\Lambda}$CDM model and XCDM model, in which the linear size of
this standard ruler was calibrated as $l_m=11.03$ pc through a
cosmological- model-independent method. As complementary to other
cosmological standard rods, such as BAO and galaxy clusters
\citep{bonamente2006determination}, quasars are promising objects
for studying the expansion rate of the Universe at much higher
redshift, thus have become an effective probe in cosmology and
astrophysics \citep{cao2017ultra,zheng2017ultra}. As an extension of
the previous work \citep{cao2017ultra}, the aim of this analysis is
to test alternative cosmological models using the quasar sample and
investigate its possibility to provide additional information of
model discrimination compared with that provided by other standard
ruler data (BAO and galaxy clusters). Two model diagnostics, the
$Om$ diagnostic \citep{sahni2008two} and the statefinder diagnostic
\citep{sahni2003statefinder}, are also applied to our work.

This paper is organized as follows: In Section~\ref{sec:dataset}, we
briefly introduce the observational data sets used. The details of
the cosmological models to be considered are presented in
Section~\ref{sec:model}. In Section~\ref{sec:Method}, we describe
the methods used to obtain the constraints for each data set. In
Section~\ref{sec:res}, we perform a Markov Chain Monte Carlo (MCMC)
analysis, and furthermore apply model diagnostics in
Section~\ref{sec:diag}. The main results are summarized in
Section~\ref{sec:con}.

\begin{figure}[ht!]
\begin{center}
\includegraphics[width=0.45\textwidth]{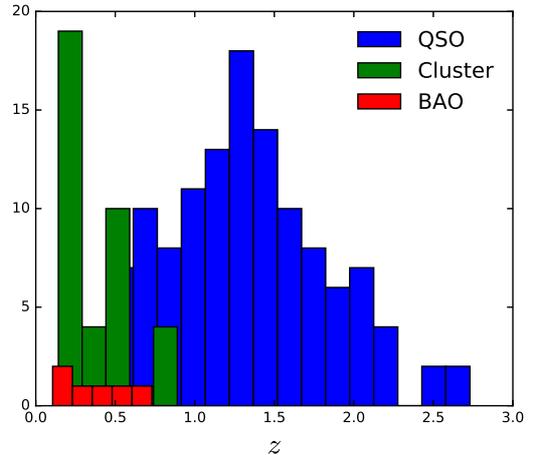}
\end{center}
\caption{Redshift distribution of different standard ruler data. One
can see a fair coverage of redshifts in the combined sample.
\label{fig:redshift}}
\end{figure}

\section{Data} \label{sec:dataset}

Three types of standard rulers currently available are used to place
constraint on different cosmological models: the compact radio
quasars (QSO) data from VLBI, the angular-diameter distance ($D_A$)
measurements derived from galaxy clusters, and the baryonic acoustic
oscillations (BAO) data from large-scale structure (LSS)
observations.

${\bf{QSO}}$.  In our analysis, we use the angular size measurements
of 120 radio quasars covering the redshift range of $0.46<z<2.76$
\citep{cao2017ultra}. The linear sizes of compact structure in
intermediate-luminosity radio quasars ($10^{27}$ W/Hz $<L<10^{28}$
W/Hz) display negligible dependent on luminosity and redshift
\citep{cao2017a}. We refer the reader to \citep{cao2017a} for the
detailed selection methodology to obtain the final sample of radio
quasars explicitly presented in \citep{cao2017ultra}, which could serve
as standard cosmological rods with intrinsic linear size calibrated
to $l_m=11.03\pm0.25$ pc.

{\bf{Galaxy cluster}}. X-ray observation of intracluster medium and
radio observations of Sunyaev-Zeldovich effect allow us to estimate
the angular diameter distance of galaxy clusters. In this paper, we
will use the $D_A$ measurements of 38 galaxy clusters in the
redshift range of $0.16<z<0.89$. The final statistical sample with
all necessary information can be found in
Refs.~\citep{bonamente2006determination}.

{\bf{BAO}}. The third astrophysical probe applied to the joint
cosmological analysis is BAO, which measures the angular-diameter
distance through the clustering of galaxies perpendicular to the
line of sight. The acoustic peak in the galaxy correlation function
has been detected over a redshift range of $0.1<z<0.7$ with large
scale surveys. The determination of the BAO scale at lower redshift,
$z=0.106$, was made in the 6dFGS survey \citep{beutler20116df},
while the other four measurements of acoustic scale at higher
redshifts were respectively obtained by SDSS-DR7
\citep{padmanabhan20122}, SDSS-DR9 \citep{anderson2012Clustering},
and the WiggleZ galaxy survey \citep{blake2012wigglez}. These data
extensively used in nine-year WMAP analysis were summarized in Table
1 of Ref.~\cite{hinshaw2013nine}.

We remark here that, in order to test the cosmological models at
higher redshifts, it is very necessary to turn to distance
indicators located in the deeper universe. Higher-redshift
radio-loud quasars are valuable additions to standard rulers used
for cosmological tests, since the predictions of cosmological models
can be radically different. Fig.~\ref{fig:redshift} shows the
redshift coverage of different standard ruler data. One can see that
inclusion of quasars could result in a fair coverage of redshifts,
which enables QSO an excellent complement to other observational
probes at lower redshifts.

\section{Cosmological models}\label{sec:model}

Four cosmological models are considered with the data sets described
above: Chevallier-Ploarski-Linder parametrization (CPL), Entropy
Dark energy model (EDE), Generalized Chaplygin Gas Model (GCG) and
modified polytropic Cardassian model (MPC). A flatness of Friedmann
Robertson Walker (FRW) \textbf{metric} is assumed in our analysis,
which is strongly supported by the recent observations
\citep{amanullah2010spectra,hinshaw2013nine}. Under this assumption,
the angular diameter distance can be expressed as
\begin{equation}{\label{equ:DDA}}
{D_A(z)}\,=\,{\frac{c}{H_0(1+z)}}\int_0^z\,\frac{dz'}{E(z')}
\end{equation}
where $E(z)\,=\,H(z)/H_0$ and $H_0$ is the Hubble constant, which is
fixed at $67.8\pm0.9\,\rm{km\,s^{-1} Mpc^{-1}}$ based on recent
\textit{Planck} results \citep{ade2016planck}. Moreover, in order to
obtain stringent constraints on other cosmological parameters, for
the CPL, EDE, and MPC models, we use the prior on the matter density
parameter $\Omega_m$ from \textit{Planck} \citep{ade2016planck}. It
should be noted that, although the above priors always influence the
cosmological analysis, conclusions should not be significantly
affected concerning the core of this paper, i.e., the comparison of
the confidence regions derived from different standard ruler data.

\subsection{CPL model}

A simple extension of the $\rm{\Lambda}$CDM model is the XCDM model
with constant equation of state. However, it would be natural to
consider the equation of state varying with redshifts, i.e., it
could be an arbitrary function of the redshift, $w\,=\,w(z)$. One of
the most popular functions is the CPL parametrization
\citep{chevallier2001accelerating, linder2003cosmic}
\begin{equation}
{w(z)}\,=\,{w_0+w_a\frac{z}{1+z}}
\end{equation}
where $w_0$ and $w_a$ are the two parameters to be fitted by the
observational data. Note that the $\Lambda$CDM model can be always
recovered by taking $w_0=-1$ and $w_a=0$. Then the Hubble function
can be expressed as
\begin{equation}\label{eq:CLP}
\begin{aligned}
{H(z)}\,=&\,H_0[\Omega_m(1+z)^3+\\
&\Omega_{DE}(1+z)^{3(1+w_0+w_a)}e^{(\frac{-3w_az}{1+z})}]^{1/2}
\end{aligned}
\end{equation}
where $\Omega_{DE}\,=\,1-\Omega_m$, and the cosmological parameters
in this cosmological model are ${\bf{p}}\,=\,(w_0, w_a)$.

\subsection{EDE model}

Recently, the Entropy Dark Energy model was proposed on the base of
the theory of entropic gravity \cite{easson2011entropic,
easson2012entropic}. In the framework of entropic gravity theory,
the gravity force can be explained as a kind of entropic force
related to the change of entropy, while the field of equation of
gravity is obtained with the second law of thermodynamics. The EDE
in the entropic gravity model arises from the surface term in the
Einstein-Hilbert's action. In the previous work of
Refs.~\cite{easson2011entropic,easson2012entropic}, a positive term
$C_HH^2+C_{\dot{H}}{\dot{H}}$ (the overdot means a derivative with
time) was added to the surface part in the action, where $C_H$ and
$C_{\dot{H}}$ are the model parameters respectively falling into the
range of $3/2\pi\leq C_H \leq 1$ and $0 \leq C_{\dot{H}} \leq
3/2\pi$. In our work, no interaction between the DE and other cosmic
components (especially matter) is assumed and one can derive the
evolution of Hubble parameter as
\begin{equation}\label{eq:EDE}
H(z)\,=\,H_0[\eta(1+z)^3+(1-\eta)(1+z)^{2(C_H-1)/C_{\dot{H}}}]^{1/2}
\end{equation}
where
\begin{equation}
\eta\,=\,\frac{\Omega_m}{1+(\frac{3}{2}C_{\dot{H}}-C_H)}
\end{equation}
It is straightforward to show that $\eta = \Omega_m$ when $C_H =
3C_{\dot{H}}/2$, and the parameters to be considered in this model
are ${\bf{p}}\,=\,(C_H, C_{\dot{H}})$.

\subsection{GCG model}

The so-called General Chaplygin Gas model (GCG) has been widely
studied to explain the accelerating universe
\citep{kamenshchik2001alternative,barreiro2008wmap,lu2009observational}.
In the GCG model, the dark sectors in the Universe, i.e., dark
energy and dark matter, can be unified through an exotic equation of
state. More specifically, the GCG background fluid with its energy
density $\rho_{GCG}$ and pressure $p_{GCG}$ can be related with the
equation of state \citep{kamenshchik2001alternative} $p_{GCG} =
-\frac{A}{\rho_{GCG}^\alpha}$, where
$\rho_{GCG}\,=\,\rho_{DE}+\rho_{DM}$ is the unified energy density
of dark energy and dark matter. The Universe is filled with two
components, the GCG component and the baryonic matter component,
i.e., $\rho = \rho_{GCG} +\rho_b$. Under the assumption of flat FRW
metric, the Hubble parameter of this model can be expressed as
\begin{equation}\label{eq:GCG}
\begin{aligned}
H(z)\,=\,&H_0[\Omega_b(1+z)^3+\\
&\Omega_{GCG}(A_s+(1-A_s)(1+z)^{3(1+\alpha)})^{\frac{1}{1+\alpha}}]^{1/2}
\end{aligned}
\end{equation}
where $\Omega_{GCG}=1-\Omega_b$ and $A_s =
\frac{A}{\rho^{1+\alpha}}$.  In our analysis, the baryonic density
parameter is fixed at $\Omega_b = 0.0484$ based on the recent
\textit{Planck} results \cite{ade2016planck}, and the two
cosmological parameters in this model are ${\bf{p}}\,=\,(A_s,
\alpha)$.

\subsection{MPC model}

In order to explain the cosmic acceleration from a different
perspective, Freese \& Lewis (2002) put forward a Cardassian model
without the introduction of dark energy
\citep{freese2002cardassian}, for which the Friedmann equation is
modified as
\begin{equation}
H^2\,=\, \frac{8\pi G \rho_m}{3}+B\rho ^n_m
\end{equation}
where $\rho_m$ is the total matter density. We emphasize here that,
in order to lead to the cosmic acceleration in this
parameterization, the value of the parameter $n$ should always be
$n<2/3$. Then a simple generalized case of the Cardassian model was
proposed in Ref.~\citep{gondolo2002accelerating}, in which an
additional exponent $q$ was introduced. The Hubble parameter with
this generalization can be written as
\begin{equation}\label{eq:MPC}
H(z)\,=\, H_0\{\Omega_m(1+z)^3 \times[1+(\Omega_m^{-q}-1)(1+z)^{3q(n-1)}]^{1/q}\}^{1/2}
\end{equation}
where the parameters to be constrained are ${\bf{p}}\,=\,(q,n)$.
This MPC model will reduce to $\rm{\Lambda}$CDM model with $q=1$ and
$n=0$.

\section{Method}\label{sec:Method}

In the following, we consider the observational constraints on the
cosmological models from observational data.

\subsection{QSO}

If taking milliarcsecond structure in radio quasars as individual
standard rulers, the angular sizes at redshift $z$ can be written as
\begin{equation}\label{eq:theta}
\theta(z)\,=\,\frac{l_m}{D_A(z)}
\end{equation}
where $D_A(z)$ is the corresponding angular diameter distance
mentioned above and $l_m$ is the intrinsic length of milliarcsecond
structure in radio quasars. In this work, we take the typical value
of $l_m=11.03\pm0.25$ pc calibrated with cosmic chronometers, which
was obtained through a new cosmology-independent calibration
technique \citep{cao2017ultra}. Following the general classification
of Active Galactic Nuclei (AGN), it is powered by the accretion of
mass onto black holes in the center of galaxies and will produce
jets of relativistic plasma in the central regions. There are two
main physical meaning related to the linear size of this standard
ruler: on the one hand, there is almost no stellar contribution when
the distance from the AGN center approaches 10 pc, which is also the
position at which AGN jets are typically generated
\citep{Blandford78}; On the other hand, according to the recent
analysis of the correlation between the black hole's mass accretion
and the star-formation rate, 10 pc represents the typical radius
within which the two rates are almost equal to each other, a
conclusion supported by the findings from both recent observations
and simulations of AGN \citep{silverman2009the,hopkins2009how}. More
importantly, the value of $l_m$ estimated from single-frequency VLBI
measurements agrees very well with that obtained from
multi-frequency VLBI imaging observations
\citep{pushkarev2015milky}. Such consistency could also be seen from
the comparison between the cosmological fits derived from two types
of VLBI observations at different observing frequencies
\citep{Caomulti}. The data points of the 120 QSOs are given in terms
of the angular sizes, $\theta_i^{obs}$. One can then constrain
cosmological parameters by minimizing the $\chi^2$ function given by
\begin{equation}
\chi_{QSO}^2\,=\,\sum_i^{120}
{\frac{(\theta(z_i;{\bf{p}})-\theta_i^{obs})^2}{\sigma_i^2}}
\end{equation}
where $\theta(z_i;\bf{p})$ is the theoretical value of the angular
size at redshift $z$ (which is defined in Eq.~(\ref{eq:theta})) and
${\bf{p}}$ represents the cosmological parameters of interest (which
are specifically introduced in Section \ref{sec:model}).
$\theta_i^{obs}$ is the observed counterpart of the angular size for
the ith quasar. Following the work of Ref.~\citep{cao2017ultra}, in
our analysis the total uncertainty expresses as
$\sigma^2_{i}\,=\,\sigma^2_{sta,i}+\sigma^2_{sys,i}$, which includes
the statistical error of observations in $\theta_i^{obs}$ and an
additional $10\%$ systematical uncertainty accounting for the
intrinsic spread in the linear size. See Table 1 of
Ref.~\citep{cao2017ultra} for details of the quasar data and
reference to the source papers.

\subsection{Galaxy clusters}

We can obtain the angular diameter distances by using the
Sunyaev-Zeldovich effect together with x-ray emission of galaxy
clusters, which can be directly used to estimate the cosmological
parameters by minimizing the corresponding $\chi^2$ as
\begin{equation}
\chi_{Cluster}^2\,=\,\sum^{38}_{i=1}\,\frac{[D_A^{th}(z_i;{\bf{p}})-D_A^{obs}(z_i)]^2}{\sigma^2_{D_A,i}}
\end{equation}
Here $D_A^{th}(z_i;{\bf{p}})$ is the theoretical angular diameter
distance at redshift $z_i$, which is defined in Eq.~(\ref{equ:DDA}).
$D_A^{obs}(z_i)$ is the observed angular diameter distance of the
$i$th galaxy cluster with total uncertainty defined as
$\sigma^2_{D_A,i} = \sigma^2_{mod}+\sigma^2_{stat}+\sigma^2_{sys}$,
where the modeling error ($\sigma_{mod}$), statistical error
($\sigma_{stat}$) and systematical error ($\sigma_{sys}$) are
explicitly shown in Table~2-3 in
Ref.~\cite{bonamente2006determination}.

\subsection{BAO}

Compared with the previous works involving BAO as a standard ruler
\citep{cao2010testing,cao2011testing,cao2011constraints}, we use the
measurement of distance ratio $r_s(z_d)/D_V(z)$ or $D_V(z)/r_s$ from
the BAO peak in the distribution of SDSS luminous red galaxies,
which contains the main information of the observations of LSS. Here
$r_s(z_d)$ is the comoving sound horizon at the drag epoch, where
the redshift $z_d$ at the baryonic drag epoch is fitted with the
formula proposed in Ref.~\cite{eisenstein1998baryonic}. $D_V(z)$ is
the effective distance given by
\begin{equation}
D_V(z)\,=\,\left[(1+z)^2D_A^2(z)\frac{cz}{H(z)}\right]^{1/3}
\end{equation}
where $D_A$ is the angular diameter distance and $H(z)$ is Hubble
parameter. The $\chi^{2}$ function for BAO is defined as
\begin{equation}
\chi^{2}_{BAO}=\bf(x-d)^{T}(C_{BAO}^{-1})(x-d),
\end{equation}
where
\begin{equation}
\begin{aligned}
{\bf{x-d}}=[r_{s}/D_{V}(0.1)-0.336, D_{V}(0.35)/r_{s}-8.88,\\
 D_{V}(0.57)/r_{s}-13.67,r_{s}/D_{V}(0.44)-0.0916,\\
r_{s}/D_{V}(0.60)-0.0726, r_{s}/D_{V}(0.73)-0.0592]
\end{aligned}
\end{equation}
and $C^{-1}_{BAO}\,=$
\begin{eqnarray*}
\left(
\begin{array}{cccccc}
4444.4   &  0       & 0         &  0          & 0         &  0        \\
0        &  34.602  & 0         &  0          & 0         &  0        \\
0        &  0       & 20.661157 &  0          & 0         &  0        \\
0        &  0       & 0         &  24532.1    & -72584.4  &  12099.1  \\
0        &  0       & 0         &  -25137.7   & 134598.4  &  -64783.9 \\
0        &  0       & 0         & 12099.1     & -64783.9  &  128837.6 \\
\end{array}\right)
\end{eqnarray*}
is the corresponding inverse covariance matrix
\citep{hinshaw2013nine}.

\subsection{Joint analysis}

We will present a combined analysis of the above three tests to fit
theoretical models to observational data. Meanwhile, in order to
make a good comparison with the quasar sample, a joint analysis with
galaxy clusters and BAO data sets is also performed in this
analysis. The $\chi^2$ function of the above two combined analysis
are respectively expressed as
\begin{equation}
\chi^2_{total}\,=\,\chi^2_{BAO}+\chi^2_{Cluster}+\chi^2_{QSO}
\end{equation}
and
\begin{equation}
\chi^2_{BAO+Cluster}\,=\,\chi^2_{BAO}+\chi^2_{Cluster}
\end{equation}

\begin{table}{}
\centering \caption{The marginalized 1$\sigma$ errors of the model
parameters for different cosmological scenarios, as well as their
corresponding $FoM$, estimated from QSO, BAO, galaxy clusters and
the joint analysis.}\label{tab:CPL}
\begin{tabular}{|c|c|cc|c}
\hline \hline
Data Set           & FoM         & $w_0$                     & $w_a$\\
\hline
QSO                  &  $27.40$    & $-0.91^{+0.48}_{-0.54} $  & $ -0.4^{+2.9}_{-2.3}$ \\
BAO                   & $1.61$      & $-1.12^{+0.31}_{-0.35} $  &  $-1.8^{+3.2}_{-2.1}$   \\
Cluster              & $29.35$     & $> -0.286$                &  $-5.7^{+1.3}_{-2.8}$ \\
Cluster+BAO          & $20.90$     & $-0.23^{+0.22}_{-0.36} $ & $-4.0^{+4.3}_{-3.4} $  \\
All                  &  $111.758$  & $-0.72^{+0.19}_{-0.32}$   & $-1.43^{+1.7}_{-0.87}$ \\
\hline \hline
Data Set      &  FoM          & $C_{\dot{H}}$                & $C_H$\\
\hline
QSO          &  $117950.7$   & $> 0.327$                    & $0.793^{+0.099}_{-0.057}$ \\
BAO            &  $62181.8$    & $ > 0.419$                   &  $0.675^{+0.14}_{-0.096}$   \\
Cluster      &  $104122.7$   & $0.367^{+0.11}_{-0.046}$     &  $> 0.875$ \\
Cluster+BAO     &  $1949008.7$  & $>0.470 $                     & $0.933^{+0.057}_{-0.022} $  \\
All          &  $3185494.1$  & $> 0.469$                    & $0.895^{+0.040}_{-0.027}$ \\
\hline \hline
Data Set     &   FoM        & $A_s$                      & $\alpha$\\
\hline
QSO           &  $5819.5$    & $0.761^{+0.080}_{-0.075}$  & $> -0.0945$ \\
BAO           &  $4424.4$    & $0.890^{+0.036}_{-0.168}$  &  $0.24\pm 0.39$   \\
Cluster       &  $963.7$     & $0.575^{+0.108}_{-0.050}$    &  $< -0.0954$ \\
Cluster+BAO   &  $64683.43$  & $0.621^{+0.082}_{-0.067} $ & $-0.25^{+0.17}_{-0.14} $  \\
All           &  $146723.6$  & $0.708\pm 0.040$           & $-0.05^{+0.10}_{-0.16}$ \\
\hline \hline
Data Set      &  FoM      & $\beta$                  & $n$\\
\hline
QSO          &  $1.99$   & $0.56^{+5.34}_{-0.36}$   &  $0.46^{+0.166}_{-0.472}$ \\
BAO           &  $1.89$   & $ 0.61^{+4.49}_{-0.41}$  &  $0.09^{+0.33}_{-0.45}$   \\
Cluster      &  $2.66$   & $0.32^{+5.5}_{-0.33}$    &  $0.56^{+0.041}_{-0.31}$ \\
Cluster+BAO     & $1.70$    & $0.38^{+3.02}_{-0.28} $  &  $0.122^{+0.216}_{-0.286} $  \\
All          &  $3.94$   & $0.59^{+3.0}_{-0.33}$    &  $0.46^{+0.031}_{-0.29}$ \\
\hline \hline

\end{tabular}
\end{table}

\section{Results and discussion} \label{sec:res}

For all cosmological models described in Section~\ref{sec:model}, we
estimate the constraint ability of different angular diameter
distance data, QSO, BAO, galaxy clusters, BAO+Cluster and
BAO+Cluster+QSO, by minimizing the $\chi^2$ function given in
Sect.~\ref{sec:Method}. Furthermore, the Figure of Merit (FoM)
\citep{mortonson2010figures} is also applied to quantify the
constraining power of each data (especially the quasar sample) on
cosmological model parameters.

\subsection{CPL}

\begin{figure}[ht!]
\begin{center}
\includegraphics[width=0.45\textwidth]{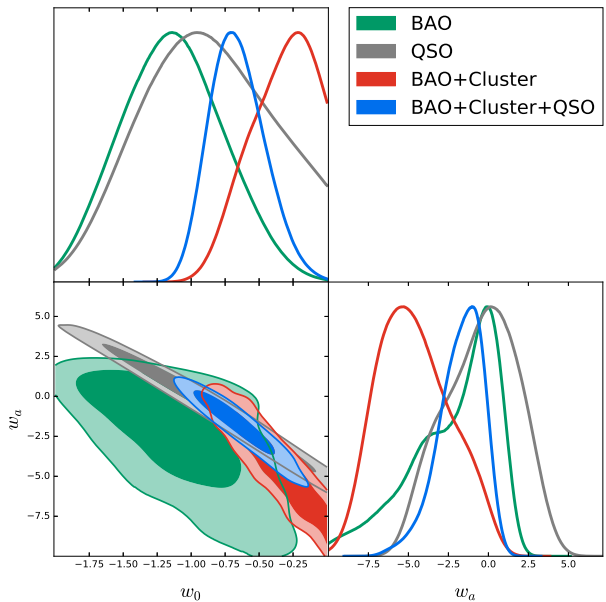}
\end{center}
\caption{$1\sigma$,$2\sigma$ contours of the CPL model parameters
$w_0$ and $w_a$ obtained from different standard ruler data.
\label{fig:CPL}}
\end{figure}

The best fits for CPL parameters ${w_0, w_a}$ and the estimated
$\chi^2$ from different data sets are shown in Table \ref{tab:CPL}.
The $1\sigma$, $2\sigma$ contours of the model parameters are
presented in Fig.\ref{fig:CPL}. As can be seen from Table
\ref{tab:CPL} and Fig.\ref{fig:CPL}, the fitting results from QSO
are in good agreement with those from BAO, whereas in tension with
the results from galaxy cluster data. Notice that concordance
$\rm{\Lambda}$CDM cosmology ($w_0=-1$, $w_a=0$) is consistent with the
quasar and BAO standard ruler data at less than 1$\sigma$ level.
More importantly, compared with the previous literature using other
independent and precise experiments
\citep{cao2011constraints,cao2012testing,cao2015cosmology}, the
currently compiled quasar data may improve the constraints on model
parameters significantly, in the framework of CPL parametrization.
When adding QSO data set to the joint data set of BAO and galaxy
cluster, we will get more precise assessment of ${w_0, w_a}$, which
is consistent with that obtained from the recent \textit{Planck} CMB
data \citep{ade2016planck} as well as the combination of the CMB
measurements from Atacama Cosmology Telescope (ACT) and the South
Pole Telescope, BAO and $H_0$ measurements \citep{hinshaw2013nine}.

\begin{figure}[ht!]
\begin{center}
\includegraphics[width=0.45\textwidth]{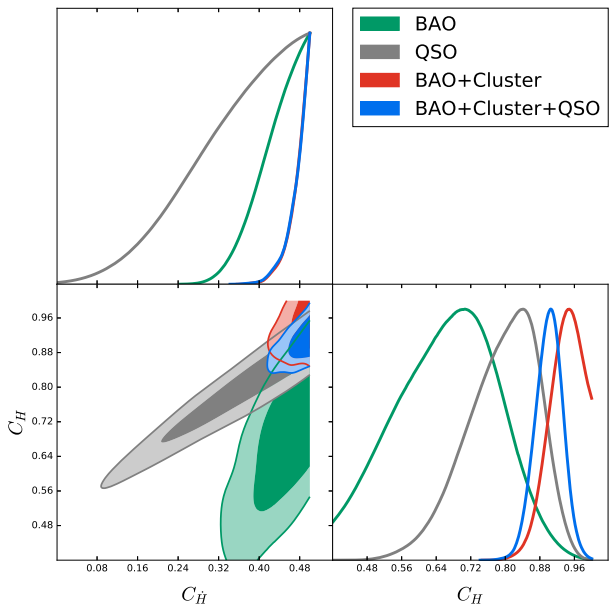}
\end{center}
\caption{$1\sigma$,$2\sigma$ contours of the EDE model parameters
$C_{\dot{H}}$ and $C_H$ obtained from different standard ruler data.
\label{fig:EDE}}
\end{figure}

\subsection{EDE}

Table~1 shows the best fits of EDE parameters $(C_{\dot{H}}, C_H)$
and the $\chi^2$ derived from different observational data sets.
Although the exact value of $C_{\dot{H}}$ are not independently
obtained with QSO or BAO, appreciable consistency between the same
type of probes (standard rulers) is indeed revealed in our analysis.
On the other hand, it is clear that the quasar data set could
provide constraints on the other parameter $C_H$ comparable to the
other two types of standard probes. These implications can be more
clearly seen from the corresponding contours for EDE model, which
are explicitly presented in Fig.~\ref{fig:EDE}. Fitting results from
the joint angular diameter distance data of QSO+BAO+Cluster give the
best-fit parameters $C_{\dot{H}} > 0.469$ and $C_H =
0.895^{+0.040}_{-0.027}$, which agree very well with the results
yielded from the luminosity distance data including 307 Type Ia
Supernovae: $C_{\dot{H}} = 0.415\pm0.061$ and $C_H = 0.813\pm0.056$
\citep{mathew2016evolution}.

\begin{figure}[ht!]
\begin{center}
\includegraphics[width=0.45\textwidth]{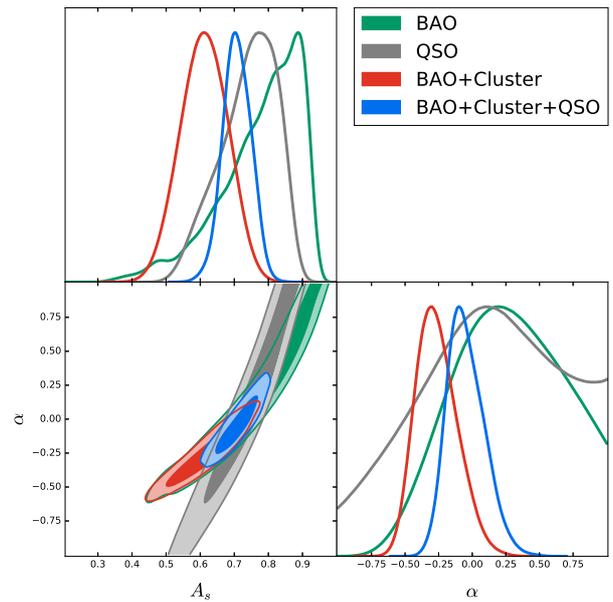}
\end{center}
\caption{$1\sigma$,$2\sigma$ contours of the GCG model parameters
$A_s$ and $\alpha$ obtained from different standard ruler data.
\label{fig:GCG}}
\end{figure}

\subsection{GCG}

Working on the GCG model, we obtain the fitting results from
different combinations of observational data, which are displayed in
Table.~1 and Fig.~\ref{fig:GCG}. The excellent consistency between
the three types of standard rulers can be clearly seen through the
comparison of these plots. The joint analysis with standard rulers
provides the best-fit parameters and the marginalized 1$\sigma$
constraints as $A_s=0.708\pm 0.040$ and
$\alpha=-0.05^{+0.10}_{-0.16}$. For comparison, it is necessary to
refer to the previous results obtained with other astrophysical
measurements. The results obtained with the combination analysis of
the X-ray mass fractions of galaxy clusters, the dimensionless
coordinate distance to SN \uppercase\expandafter{\romannumeral1} and
FR\uppercase\expandafter{\romannumeral2}b radio galaxies gave
$A_s\,=\,0.70^{+0.16}_{-0.17}$ and
$\alpha\,=\,-0.09^{+0.54}_{-0.33}$ \citep{zhu2004generalized}. The
former work done in Ref.~ \citep{wu2007generalized} with the joint
data of 9 Hubble parameters data points, 115 SN
\uppercase\expandafter{\romannumeral1}a and BAO peak at $z\,=\,0.35$
showed that $0.67 \le A_s \le 0.83$ and $-0.21\le \alpha \le 0.42$.
Recent work done in \cite{lu2009observational} indicated
$A_s\,=\,0.73\pm 0.06$ and $\alpha\,=\,-0.09^{+0.15}_{-0.15}$, which
strengthens the indication that joint analysis of cosmic standard
rulers (QSO+BAO+Cluster) could provide consistent but more stringent
fitting results compared with these previous results.

\begin{figure}[ht!]
\begin{center}
\includegraphics[width=0.45\textwidth]{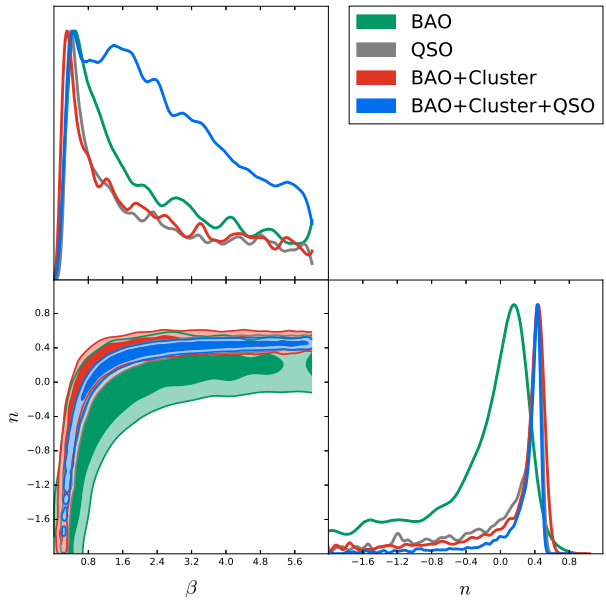}
\end{center}
\caption{$1\sigma$,$2\sigma$ contours of the MPC model parameters
$\beta$ and $n$ obtained from different standard ruler data.
\label{fig:MPC}}
\end{figure}

\subsection{MPC}

All of the fitting results obtained with QSO, BAO, Cluster and the
joint data are presented in Table~1. The $1\sigma$, $2\sigma$
contours of the MPC model parameters $\beta$ and $n$ are also
illustrated in Fig.~\ref{fig:MPC}. The results from QSO, BAO and
galaxy cluster are consistent with each other within $1\sigma$
confidence level. Several authors have tested the MPC model using
various data sets. For instance, Cosmic All-Sky Survey (CLASS)
lensing sample \citep{alcaniz2005constraints} gave $\beta\,=\,0.05$,
$n\,=\,-2.32$, which is in tension with our results from
QSO+BAO+Cluster. However, It is important to note that the shape of
the constraint contours derived in our analysis are very similar to
those shown in
Refs.~\citep{alcaniz2005constraints,feng2010cardassian,liang2011latest,li2011testing,magana2015magnified}.
Moreover, the results obtained with standard rulers turned out to
correspond well with previous works. Our results are similar to the
results obtained with SN Ia+BAO+WMAP5
($\beta\,=\,0.48^{+2.020}_{-0.080}$,
$n\,=\,-0.600^{+0.980}_{-0.450}$) and SN Ia+BAO+CMB data sets
($\beta\,=\,1.098^{+1.015}_{-0.465}$,
$n\,=\,-0.041^{+0.364}_{-0.964}$) at $1\sigma$ confidence level
\citep{feng2010cardassian}. More recent works
\citep{magana2015magnified} have suggested the cutoffs of
$0.45\,<\beta\,<1.05$ and $-0.8\,<n\,<0.05$, which achieved a
similar precision to our work.

\begin{figure}[ht!]{}
\begin{center}
\includegraphics[width=0.5\textwidth]{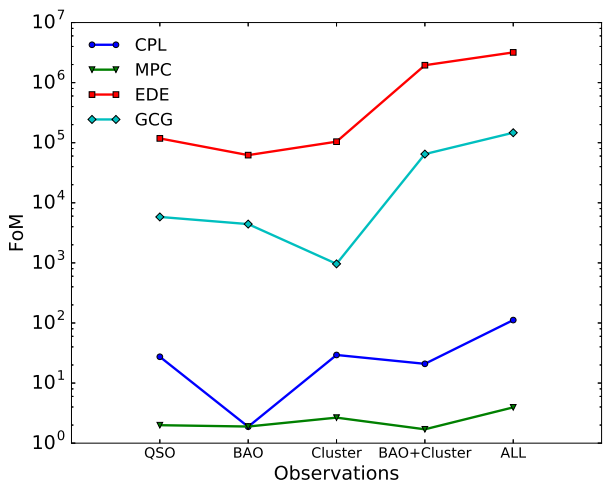}
\end{center}
\caption{FoM for different cosmological models using QSO, BAO,
galaxy cluster observations and the joint data sets.
\label{fig:FoM}}
\end{figure}

\subsection{Figure of Merit}

As discussed above, the method of using the angular size
measurements of compact structure in radio quasars distance could
provide a complementary and effective probe in cosmological
applications. However, in order to quantify the constraining power
of the quasar sample, we introduce the Figure of Merit (FoM)
\citep{albrecht2006report, mortonson2010figures}, a useful
statistical tool originally defined by the Dark Energy Task Force as
the inverse of the area enclosed by the $95\%$ confidence level
contour of CPL parameters, $(w_0, w_a)$. Later there was a more
general definition of FoM \citep{wang2008figure}
\begin{equation}
\text{FoM}\,=\,\frac{1}{\sqrt{\text{det\,Cov}(f_1, f_2, f_3,...)}}
\end{equation}
where $\text{Cov}(f_1,f_2,f_3,...)$ is the covariance matrix of the
cosmological parameters $f_i$. Note that a larger FoM corresponds to
a smaller error ellipse, which therefore denotes a tighter
constraints on the cosmological parameters.

We have calculated the FoM of the cosmological models for each data
set analyzed, which is explicitly summarized in Table~1. A graphical
representation of the FoM results is also provided in
Fig.~\ref{fig:FoM}, which directly shows the results in the FoM test
for each cosmological model. Out of all the candidate models
considered, it is obvious that the QSO data could provide better
statistically constraints on cosmological parameters than BAO. This
could attribute to the large sample size and the higher redshift
range covered by QSOs in comparison to other cosmological probes.
When taking galaxy cluster observations into consideration, quasars
perform better than galaxy clusters in the framework of two
cosmological models, GCG and EDE models. On the other hand, when
comparing the FoM of BAO+Cluster and that of all observations, we
find the inclusion of the QSO sample will generate more stringent
cosmological constraints.

\section{Model diagnostics}\label{sec:diag}

In order to discriminate the four cosmological scenarios considered
in this paper, it is important to find sensitive and robust
diagnostics to illustrate the dynamic behavior of different
cosmologies. As is well known, the expansion rate of the Universe
can be expressed by the Hubble parameter $H\,=\,\dot{a}/a$, where
$a$ is the scale factor, while the rate of cosmic acceleration ia
always quantified by the deceleration parameter
\begin{equation}
q\,=\,-\frac{\ddot{a}}{aH^2}\,=\,-\frac{a\ddot{a}}{{\dot{a}}^2}
\end{equation}
However, it is very difficult for the Hubble parameter $H$ and the
deceleration parameter $q$ to accurately distinguish cosmological
models cause all the models considered will give similar results, e.g., $\ddot{a}\,>\,0$ and $H\,>\,0$
or $q\,<\,0$, which encourages us to invoke some newer and more
effective quantities to substitute the two original parameters. In
this work, we will take the $Om$ diagnostic and the statefinder
diagnostic into consideration.

\begin{figure}[ht!]{}
\begin{center}
\includegraphics[width=0.5\textwidth]{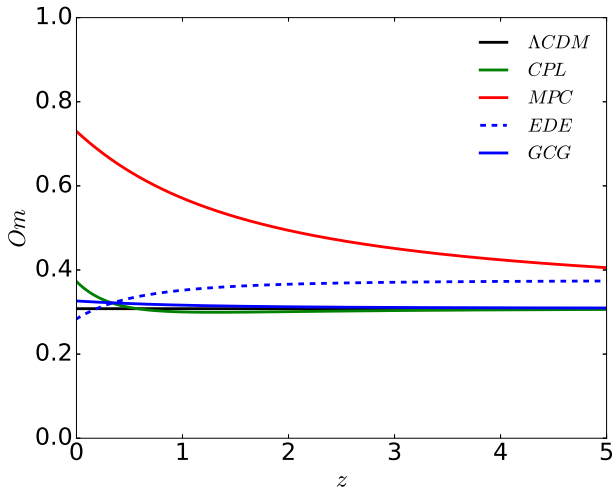}
\end{center}
\caption{ The evolution of $Om(z)$ versus the redshift $z$ for
different cosmological models. \label{fig:Omz}}
\end{figure}


It is well known that $Om(z)$ is a combination of the Hubble
parameter and the redshift, which provides a null test of dark
energy being a cosmological constant at different stages for the
$\rm{\Lambda}$CDM model \citep{sahni2008two}. Therefore, this
diagnostic, which has been extensively used to discriminate
different cosmological models as well as $\rm{\Lambda}$CDM model
\citep{ding2015is,zheng2016what}, can be defined as
 \begin{equation}\label{equ:Om}
Om(z)\,=\, \frac{E^2{(z)}-1}{(1+z)^3-1}
 \end{equation}
where $E(z)\,=\,H(z)/H_0$. In the basic $\rm{\Lambda}$CDM model
neglecting the radiation at low redshifts, one can easily get
\begin{equation}\label{equ:ELCDM}
E^2_{\rm{\Lambda} CDM}(z) = \Omega_m(1+z)^3+(1-\Omega_m)
\end{equation}
The combination of Eq.~(\ref{equ:Om}) and Eq.~(\ref{equ:ELCDM}) will
lead to
\begin{equation}
Om(z)\,|_{\rm{\Lambda} CDM}=\,\Omega_m
\end{equation}
It is obvious that $Om(z)$ should be constant and exactly equal to
the present mass density parameter $\Omega_m$ if the $\rm{\Lambda}$CDM model is the
true one, while for other cosmological models, the $Om(z)$
diagnostic are expected to give different values.

Applying the $Om(z)$ diagnostic to the cosmological models
considered in our work, we can get the relation between the redshift
and $Om(z)$ for different cosmological models, which is specifically
presented in Fig.~\ref{fig:Omz}. The $Om(z)$ for the CPL and GCG
models cannot be distinguished from each other as well as from
$\rm{\Lambda}$CDM model. Moreover, the $Om(z)$ of CPL, GCG, EDE models
cannot be distinguished at present time unless high-precision
observations are obtained and applied. Another impressive feature of
Fig.\ref{fig:Omz} is that the $Om(z)$ for the MPC model, a
cosmological candidate proposed without introducing dark energy in
the Universe, absolutely deviates from the $\rm{\Lambda}$CDM model and
other cosmological models.

\begin{figure*}[htbp]
  \begin{center}
   \resizebox{0.90\textwidth}{!}{\includegraphics{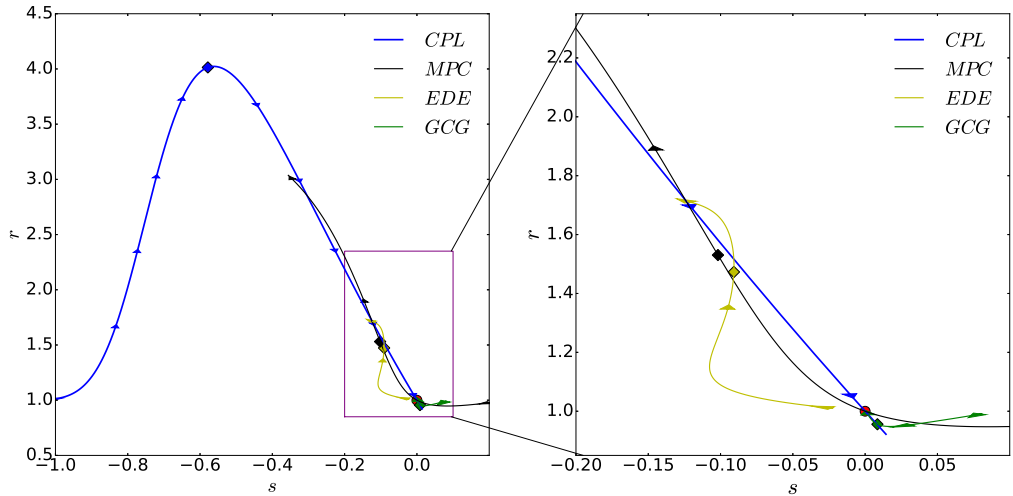}}
\caption{The evolution of the statefinder pair $(r,s)$ for various
cosmological models. The red point at $(r,s)\,=\,(0,1)$ represents
$\rm{\Lambda}$CDM model and the diamond point on each curve means the
present value of the statefinder pair $(r,s)$ for each cosmological
model. \label{fig:rs}}
  \end{center}
\end{figure*}

\begin{figure}[ht!]{}
\begin{center}
\includegraphics[width=0.5\textwidth]{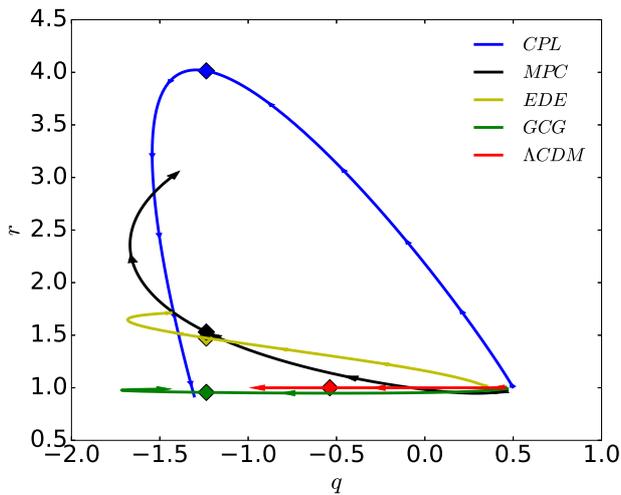}
\end{center}
\caption{The evolution of the statefinder pair $(q,s)$ for the four
cosmological models. The diamond point on each curve denotes the
current value of $(q,s)$ for each cosmological model.
\label{fig:rq}}
\end{figure}

Apart from the $Om(z)$, the statefinder diagnostic, which has been
extensively applied to discriminate different cosmological models,
involves the third derivative of the scale factor $a$ as
\citep{sahni2003statefinder}
\begin{equation}
r\,=\,\frac{\dddot{a}}{aH^3},\,\,\,\,\,\,s\,=\,\frac{r-1}{3(q-1/2)}
\end{equation}
and one can plot the corresponding trajectories in the $r-s$ plane.
For a certain cosmological model, the statefinder can be easily
derived as
\begin{equation}
r(z)\,=\,1-2\frac{E'(z)}{E(z)}(1+z)+[\frac{E''(z)}{E(z)}+(\frac{E'(z)}{E(z)})^2](1+z)^2
\end{equation}
and
\begin{equation}
s(z)\,=\,\frac{r(z)-1}{3(q(z)-1/2)}
\end{equation}
where $E(z)\,=\,H(z)/H_0$ and the deceleration parameter $q(z)$ can
be expressed as
\begin{equation}
q(z)\,=\,\frac{E'(z)}{E(z)}(1+z)-1.
\end{equation}
Applying the best fits from the joint analysis to each cosmological
model, we obtain the evolution of the statefinder $(r,s)$ and the
deceleration parameter $q$.

The evolution of the statefinder pair $(r,s)$ for different
cosmological models is shown in Fig.~\ref{fig:rs}. The red point at
$(r,s)\,=\,(0,1)$ indicates the statefinder of $\rm{\Lambda}$CDM model
and the diamond on each curve shows the present value of the
statefinder $(r,s)$ for each cosmological model. It is apparent that
the CPL model can be distinguished from other cosmological models at
present time, however, it will approach $\rm{\Lambda}$CDM in the near
future. Meanwhile, the MPC and EDE models, which are not
distinguishable from each other by the statefinder, is deviating
from $\rm{\Lambda}$CDM at the present epoch. More importantly, the GCG
model exhibits very similar evolution tendency to the concordance
cosmological constant.

The evolution trajectories in the $r-q$ plane are plotted in
Fig.~\ref{fig:rq}. For the four cosmological models considered in
this analysis, we observe the signature flip from positive to
negative in the value of $q$, which successfully explains the recent
phase transition of these models. The diamond points on different
curves in Fig.~\ref{fig:rq} denote the value of $q$ and $r$ at
present time for different cosmological models. One can see the
value of deceleration parameter $q$ is very close to each other at
present time, which is quite different from the behavior of $r$. As
for the evolution of cosmological models, at the present epoch, the
GCG model and $\rm{\Lambda}$CDM model are not distinguishable and the MPC
model can not be distinguished from EDE model. However, in the near
future they will evolve diversely, which is in well consistent with
the results obtained from the $r-s$ plot.

\section{Conclusions} \label{sec:con}

In this paper, we place constraints on four alternative cosmological
models under the assumption of the spatial flatness of the Universe:
Chevallier-Ploarski-Linder parametrization (CPL), Entropy Dark
Energy model (EDE), General Chaplygin Gas model (GCG) and Modified
Polytropic Cardassian model (MPC). A new compilation of 120
angular-size/redshift data compact radio quasars observed by
very-long-baseline interferometry (VLBI), whose statistical linear
sizes show negligible dependence on redshifts and intrinsic
luminosity and thus represents standard rulers in cosmology, are
used to test these cosmological models. Compared with BAO and galaxy
clusters acting as cosmological standard rulers, higher-redshift
radio-loud quasars are valuable additions to standard rulers used
for cosmological tests and the inclusion of quasars could result in
a fair coverage of redshifts, which enable QSO to be an excellent
complement to other observational probes at lower redshifts.

Our results show that the constraints on CPL obtained from the
quasar sample are well consistent with that obtained from BAO but in
tension with that from galaxy clusters. Note that the concordance
$\rm{\Lambda}$CDM cosmology ($w_0=-1$, $w_a=0$) is consistent with
the quasar and BAO standard ruler data at less than 1$\sigma$ level.
For other cosmological models considered, quasars provide fits in
agreement with those obtained with other probes at $1\sigma$ level.
Meanwhile, we have calculated the Figure of Merit for each
cosmological model, which is explicitly summarized in Table~1. Out
of all the candidate models considered, it is obvious that the QSO
data could provide better statistically constraints on cosmological
parameters than BAO. When taking the observations of galaxy cluster
into consideration, quasars perform better than galaxy clusters in
the framework of two cosmological models, GCG and EDE models. Based
on the best-fits obtained with QSO+BAO+Cluster, we apply two model
diagnostics, $Om(z)$ and statefinder to differentiate the dynamical
behavior of the four cosmological models. On the one hand, the
results from the $Om(z)$ diagnostic show that the CPL, GCG, EDE
models cannot be distinguished at the present epoch. However, the
MPC model, a cosmological candidate proposed without introducing
dark energy in the Universe, absolutely deviates from the
$\rm{\Lambda}$CDM model and other cosmological models. On the other
hand, in the framework of statefinder diagnostics, MPC and EDE are
will deviate from $\rm{\Lambda}$CDM model in the near future, while
GCG model cannot be distinguished from $\rm{\Lambda}$CDM model
unless much higher precision observations are available.

\section{acknowledgments}

This work was supported by the National Key Research and Development
Program of China under Grants No. 2017YFA0402603; the Ministry of
Science and Technology National Basic Science Program (Project 973)
under Grants No. 2014CB845806; the National Natural Science
Foundation of China under Grants Nos. 11503001, 11373014, and
11690023; the Fundamental Research Funds for the Central
Universities and Scientific Research Foundation of Beijing Normal
University; China Postdoctoral Science Foundation under grant No.
2015T80052; and the Opening Project of Key Laboratory of
Computational Astrophysics, National Astronomical Observatories,
Chinese Academy of Sciences. X.L. was supported by the China
Scholarship Council. This research was also partly supported by the
Poland-China Scientific \& Technological Cooperation Committee
Project No. 35-4. M.B. was supported by Foreign Talent Introducing
Project and Special Fund Support of Foreign Knowledge Introducing
Project in China.

\bibliographystyle{unsrt}

\bibliography{QSO}


\end{document}